# Dark Energy: back to Newton?

Lucy Calder and Ofer Lahav
University College London
29 January 2008

Dark Energy is currently one of the biggest mysteries in science. In this article the origin of the concept is traced as far back as Newton and Hooke in the seventeenth century. Newton considered, along with the inverse square law, a force of attraction that varies linearly with distance. A direct link can be made between this term and Einstein's cosmological constant, $\Lambda$, and this leads to a possible relation between $\Lambda$ and the total mass of the universe. Mach's influence on Einstein is discussed and the convoluted history of $\Lambda$ throughout the last ninety years is coherently presented.

## Dark Energy

The term 'Dark Energy' was coined as recently as 1999, but it is now one of the most mysterious and frequently debated topics in physics. Millions of pounds and countless hours are being spent on imaging and spectrographic surveys that aim to explain why observations point to the accelerating expansion of the universe. And with good reason. It has become apparent that the Earth, the planets, the stars and everything we are familiar with make up only a tiny 4% of the total matter and energy in the universe. Increasingly dependable evidence from gravitational lensing, galaxy rotation curves and studies of the cosmic microwave background radiation (CMB) indicates that non-baryonic Cold Dark Matter (CDM) makes up about 21% of the rest, but the remaining 75%, the Dark Energy, has not been satisfactorily explained.

It can be incorporated into Albert Einstein's general relativity (GR) by re-admitting into the field equations the cosmological constant, $\Lambda$, a term which Einstein introduced in 1917 and subsequently abandoned. In this context Dark Energy can be described as a fluid with constant energy density $\rho_\Lambda$ and pressure $p_\Lambda$, characterized by an equation of state $w = p_\Lambda / \rho_\Lambda = -1$. (Throughout this article we set $c = h = 1$.) The negative pressure gives rise to repulsive gravity and can thus accelerate the expansion of the universe once it starts dominating over matter and radiation. Current data, from type Ia supernovae, baryon acoustic oscillations and the CMB, all points to this conclusion - a flat universe with a cosmological constant - and the $\Lambda$CDM model is now standard.

However, it is also possible that the energy density changes over time and is described by a gradually evolving scalar field (in some circles the term Dark Energy is reserved for this situation). Another idea is that the cosmic acceleration is caused by inhomogeneities or by topological defects such as strings or domain walls. Even more radical is the possibility that GR is incomplete and we must develop a modified theory of gravity to account for the observations.

Since the 1960s, the cosmological constant has usually been interpreted as the zero point energy of the vacuum, similar to the ground state energy of a harmonic oscillator in quantum physics. Unfortunately, observations suggest a value for the energy density that is as much as 120 orders of magnitude smaller than that predicted by particle physics. This enormous discrepancy can be reduced to 60 orders of magnitude if the energy cutoff is taken to be of order the supersymmetric breaking scale but there is still a huge intractable difference between theory and experiment. This was articulated as the 'cosmological constant problem' by Steven Weinberg in 1989 [1].

A second cosmological constant problem became obvious in the late 1990s when the High-Z Supernova Team and the Supernova Cosmology Project both published observations on the apparent luminosity of type Ia supernovae, which indicated that the expansion of the universe is accelerating. The simplest interpretation, the ΛCDM model, requires that the vacuum energy density $\rho_\Lambda$ is presently of the same order of magnitude as the mass energy density $\rho_M$, whereas at early times (after inflation) the relative cosmological constant would have been negligible. It seems bizarre, and suggestive, that life on Earth has evolved during this brief cosmological era. Why does the epoch of galaxy formation coincide with the time when Λ starts to dominate? This requires an explanation even in models involving a dynamical Dark Energy component.

Table 1 presents the main events in the history of Dark Energy since 1917. The story is one of changing cultural attitudes, theoretical innovation and incredible technological advances. It encompasses large areas of physics and astronomy and provides a fascinating insight into the unpredictable development of science. Einstein introduced Λ in order to fit the data to his conception of the universe, but not long after he regarded it as superfluous and unattractive. Nevertheless, the addition of Λ to the gravitational field equations is the only conceivable modification that does not vastly alter the structure of the theory and this was reason enough for it not to be forgotten, while a series of observations throughout the twentieth century hinted at its presence. The number of papers on the subject markedly increased in the 1980s, when the concept of inflation arose out of particle physics and there was significant observational evidence for non baryonic dark matter; and the literature has continued to expand dramatically as the cosmological tests for the fundamental parameters have improved.

Before the publication of the supernovae data, most scientists thought it likely that, due to some as yet undiscovered mechanism, Λ was exactly zero. Aesthetically, zero seemed a much more attractive idea than almost-zero. When it became apparent that the expansion of the universe is accelerating, the cosmological constant problems suddenly assumed central importance. The strangeness of the situation points to some profound lack of understanding in fundamental physics, which may only be resolved when we have a more complete theory of quantum gravity.

**Table 1.** Dark Energy since 1917.

| Feb 1917 | Einstein adds a 'cosmological term' to his field equations. He does so because a) this allows a static universe solution to his equations and b) he believes that by doing so he guarantees Mach's principle is not violated. Einstein does not consider the cosmological constant to be part of the stress-energy term and his form for the field equations is $G_{\mu\nu} - \Lambda g_{\mu\nu} = 8\pi G T_{\mu\nu}$. This suggests that Λ is a property of space itself. | $\Lambda = 4\pi G \rho_\Lambda$ where $\rho_M$ is mass density. |
|---|---|---|
| March 1917 | De Sitter finds an apparently static solution to the modified field equations with $\rho_M = 0$, i.e. zero matter, so $\Lambda g_{\mu\nu}$ does not prevent the occurrence of inertia relative to space. The principle of | $\Lambda = 3H^2$ where $H$ is the Hubble |
| | relativity of inertia is undermined. | parameter |
| 1912-1917 | Slipher painstakingly records the spectra of 25 galaxies (then mysterious 'spiral nebulae') and finds that all but four of them are redshifted, i.e they are receding. The receding galaxies are concentrated toward Virgo and he thinks that in the opposite direction the spiral nebulae may be found to be approaching. | |
| 1922 | Friedmann shows that the field equations *without* $\Lambda$ admit nonstatic solutions with isotropic, homogeneous matter distributions corresponding to an expanding universe. However, the significance of this paper is largely ignored until 1930. | $\Lambda = 0$ |
| 1923 | Weyl points out that de Sitter's model would exhibit a redshift such as Slipher observed, increasing with distance, because although the metric in de Sitter's coordinate system is time independent, test bodies are not at rest. He and Eddington find that test particles recede from each other in the de Sitter universe. | |
| May 1923 | Einstein writes in a letter to Weyl: "If there is no quasi-static world, then away with the cosmological term." | |
| 1924 | Hubble finds faint Cepheid variables in the Andromeda nebula and realises that spiral nebulae are distant galaxies, i.e. clusters of stars far outside our own galaxy. | |
| 1927 | Lemaitre [2] makes the connection between Slipher's redshifts and a homogeneous, matter filled, expanding relativistic model. Like Friedmann's work, this paper is largely overlooked by a scientific community that still believes (due to the observed small relative velocities of the stars) the universe is static. | $\Lambda > 0$ |
| 1929 | Hubble and Humason publish a claim that the radial velocities of galaxies are proportional to their distance, i.e. the universe is expanding. In fact, Hubble originally interprets his data in the framework of the de Sitter model. | |
| 1930 | Eddington points out that Einstein's $\Lambda > 0$ static universe is unstable, although he later recognizes that Lemaitre had already shown this in 1927. "Einstein's universe is delicately poised so that the slightest disturbance will cause it to topple into a state of ever-increasing expansion or of ever-increasing contraction" [3]. Einstein has mistaken equilibrium for stability.<br><br>Unlike Einstein, Eddington believes that $\Lambda$ is an essential and irremovable foundation for cosmology because it provides a length scale against which all other lengths can be measured. In response to the instablility he develops the Eddington-Lemaitre universe, a cosmological model in which a positive $\Lambda$ allows an arbitrary long initial phase that is identical to the Einstein static universe, after which the universe begins to expand. | |
| 1931 | Einstein formally abandons $\Lambda$, calling it "theoretically unsatisfactory anyway" [4]. Gamow recalls that "when I was discussing cosmological problems with Einstein, he remarked that the introduction of the cosmological term was the biggest blunder he ever made in his life" [5]. It is debateable whether Einstein really thought this, as he never makes such a strong statement in his published work and letters. | |
| 1932 | Einstein-de Sitter model: $p \approx 0$, $\Omega_k = 0$, $\rho_M = \rho_c = 3H^2/8\pi G$. This predicts that the universe exploded from a highly condensed state and will expand monotonically forever, but at an ever decreasing rate. The density parameter $\Omega_i$ is the ratio of the density to the critical density: $\Omega_i = \rho_i/\rho_c$. | $\Omega_\Lambda = 0$<br>$\Omega_M = 1$ |
| 1933 | Lemaitre [6] identifies $\Lambda$ as equivalent to a fluid with pressure $p$ and energy density $-\rho_\Lambda$. | $p = -\rho_\Lambda$<br>$\Lambda = 4\pi G \rho_\Lambda$ |
| 1934 | McCrea and Milne [7] reveal a close correspondence between Newtonian dynamics and Einstein's GR, with the scale factor of the expansion satisfying the same equation in both theories, so long as pressure is negligible. | |
| Mid 1930s-1950s | After Einstein rejects $\Lambda$ other cosmologists retain it. One major reason is that at the time the Hubble parameter is thought to be $H_0 = 500 kms^{-1} Mpc^{-1}$, giving a corresponding age of the universe of less than 2 billion years. This made the universe younger than the geologists' predicted age of the Earth.<br><br>The Lemaitre 'hesitation universe' model is revived in which space is positively curved ($k = +1$) and more matter is present than in a static Einstein model. If $k = 1$ there is a critical value of $\Lambda$, $\Lambda_c$, such that $\dot{a} = 0$ and $\ddot{a} = 0$ can both be satisfied simultaneously. One feature of this model is that its age can greatly exceed $H_0^{-1}$. | $\Lambda = \Lambda_c(1+\varepsilon)$,<br>$\varepsilon \ll 1$ |
| 1948 | Casimir [8] shows that quantum vacuum fluctuations can produce measurable effects and calculates a force per unit area acting between two conducting plates.<br><br>Bondi, Gold [9] and Hoyle [10] develop a steady state cosmology, partly motivated by the age problem. In place of $\Lambda$, Hoyle introduces a $C$ term into the field equations, which allows a universe similar to the de Sitter model (1917) except that $\rho$ is non zero. The discovery of the CMB in 1965 is a final blow to this theory, indicating that the universe was much hotter in the past. | $\rho_\Lambda > 0$<br>$\rho_M \approx 0$ |
| 1951 | McCrea [11] shows that the steady state theory doesn't need to be viewed as a new theory of gravity with a completely new form of | |
| | matter in it. He shows that Hoyle's $C$ term is just a perfect fluid with an appropriate equation of state and so can be accommodated on the RHS of the Einstein field equations as part of the stress tensor of general relativity. | |
| 1952 | Baade [12] discovers that there are two types of Cepheid variable with different period – luminosity relations and this leads to a revised Hubble parameter of only $200 \, kms^{-1}Mpc^{-1}$. The Hubble distance scale is increased by a factor of about 2.6 and the $\Lambda$ term again becomes unnecessary. | $\Lambda = 0$ |
| 1967 | Petrosian et al [13] and others revive $\Lambda$ to explain why quasars appear to have redshifts concentrated near the value $z = 2$. In the Lemaitre loitering model, radiation emitted during the hesitation era would be received by us with almost the same redshift, so quasars born at this time could be at greatly different distances. However, with more observational evidence this explanation is thought to be inadequate. | $\Lambda = \Lambda_c(1+\varepsilon)$ $\varepsilon \ll 1$ |
| 1968 | Zeldovich [14] relates the cosmological constant to vacuum energy density and shows that quantum vacuum fluctuations must have a Lorentz invariant form $p_{vac} = -\rho_{vac}$. So the vacuum energy-momentum tensor has the form $T_{\mu\nu} = \rho_\Lambda g_{\mu\nu}$. This means that the cosmological constant contributes a term $\rho_\Lambda = \Lambda/8\pi G$ to the total effective vacuum energy. This interpretation is not new, but Zeldovich's paper is the first to convince the majority of the scientific community. (Nernst, in 1916, and Pauli in the 1920s wrote down expressions for the zero-point energy of the vacuum and wondered whether it would be gravitationally effective.) The major problem is that the various unrelated contributions to the vacuum energy predict a value up to 120 orders of magnitude greater than the observational bound! | $\rho_\Lambda(theory)$ $\approx 2 \times 10^{71}$ $GeV^4$ $\rho_\Lambda(obs)$ $\approx 10^{-47}$ $GeV^4$ |
| 1975 | Gunn and Tinsley [15] propose a non zero cosmological constant on the basis of a Hubble diagram of elliptical galaxies extending to redshifts of $z \sim 0.5$, but they acknowledge a large number of uncertainties in their conclusions. | $\Lambda > 0$ |
| Late 1970s | Electroweak theory of Weinberg, Salam and Glashow is accepted and observationally verified. This boosts interest in the cosmological constant problem. The electroweak theory is an example of a gauge theory with spontaneous symmetry breaking. | |
| 1980 | The development of spontaneously broken gauge theories and the standard model of particle physics in the 1970s leads Alan Guth and others to the concept of inflation, which can explain several cosmological problems, including why apparently causally unconnected parts of space look so similar. If the early universe | Inflation - huge early value of $\Lambda$-like |
| | were dominated by the energy density of a relatively flat real scalar field (inflaton) potential $V(\phi)$ that acts like $\Lambda$, the particle horizon could spread beyond the observable universe. This would mean that light from opposite regions of the sky was once in thermal equilibrium, which could explain the observed large scale homogeneity. | term. $a \propto e^{\sqrt{\frac{\Lambda}{3}}t}$ |
| 1982 | Guth's model is modified by Linde, Albrecht and Steinhardt, and quickly gains acceptance. The theory holds that inflation blew up quantum fluctuations in energy density from subatomic to cosmic size. This event produced the slightly inhomogeneous distribution of matter that led to the variations seen in the CMB and to the observed structures in the universe today. | |
| Early 1980s | Classification of type Ia supernovae. When their spectra are studied in detail it gradually becomes apparent they are amazingly uniform and would make excellent standard candles. | |
| 1984 | Peebles [16] and Turner, Steigman and Krauss [17] show that inflation implies vanishing curvature, $\Omega_k = 0$.<br><br>Blumenthal et al. [18] argue that the hypothesis that Dark Matter is cold (i.e. has negligible thermal velocity with respect to the Hubble flow) provides the best fit to current observations. | $\Omega_M + \Omega_\Lambda = 1$ |
| c1985 | Standard cosmology is the Einstein - de Sitter model, but the high mass density of this model is not backed up by observation unless the mass is more smoothly distributed than the visible matter.<br><br>Kaiser [19] and Davis et al. [20] show that the biased distribution of visible galaxies relative to the distribution of all of the mass can follow in a natural way in the Cold Dark Matter (CDM) theory.<br><br>Another alternative is Mixed Dark Matter (MDM), in which Cold Dark Matter is mixed with Hot Dark Matter (massive neutrinos). | $\Omega_\Lambda = 0$ $\Omega_M = 1$ $h = 0.5$ |
| 1986-1989 | Barrow and Tipler [21] and Weinberg [1] determine that the *anthropic principle* limits $\Lambda$ to a value small enough to allow the formation of sufficiently large gravitational condensations to enable life to form. "The anthropic principle has it that the world is the way it is, at least in part, because otherwise there would be no one to ask why it is the way it is" [1]. | $\rho_\Lambda \leq$ $(1+z_{max})^3$ $\times \rho_{M_0}$ |
| 1990 | Efstathiou et al. [22] show that the standard CDM model, with $\Omega_M = 1$, a Harrison-Zeldovich primordial fluctuations spectrum and a simple prescription for biasing, predicts substantially less large-scale structure than galaxy observations indicate. A remedy is to go to a universe with small $\Omega_M$, either with $\Lambda = 0$ and $\Omega_k < 0$ (spatially open) or $\Omega_k = 0$ and a non-zero cosmological constant. | $\Omega_M \approx 0.2$ $\Omega_\Lambda =$ $1 - \Omega_M$ $\approx 0.8$ |
| | The latter case becomes known as ΛCDM. | |
| 1993-1995 | It is becoming clear that fundamental observables, from the age of the universe, to the baryon content in galaxy clusters (White et al. [23]), and the nature of large-scale structure, all independently point to $\Omega_M \approx 0.2-0.3$. The case for adding a non-zero cosmological constant is becoming stronger (Ostriker and Steinhardt [24]). | $\Omega_M \approx 0.2$ to $0.3$ |
| 1997 | Martel, Shapiro and Weinberg [25] use anthropic reasoning and Bayesian probability to estimate the value of the Dark Energy density. The observed value turns out to be within their probability distribution, but they must assume that our universe is only one subuniverse in a multiverse. | $\rho_\Lambda = 1.5\rho_M$ to $2.3\rho_M$ |
| 1998-1999 | Two teams studying type Ia supernovae publish data implying an accelerating cosmic expansion [26], [27]. The simplest model to explain this is the reintroduction of Λ to the field equations. Assuming a flat universe, the best fit implies that, in the present epoch, the vacuum energy density $\rho_\Lambda$ is larger than the mass density. | $\Omega_\Lambda = \dfrac{\rho_\Lambda}{\rho_c}$ $\approx 0.7$ $\Omega_M \approx 0.3$ |
| 1999 | Turner coins the term 'Dark Energy' to describe the density $\rho_\Lambda$ that manifests itself as an effective version of Einstein's cosmological constant, but one that may vary slowly with time and position [28]. It is thought that Dark Energy might be described by a scalar field, slowly rolling toward zero over a very long time. | The equation of state: $p_i = w\rho_i$ |
| 2003 to 2006 | The conclusion that most of the energy density of the universe is a vacuum energy like the cosmological constant, causing an accelerating expansion of the universe, is greatly strengthened by temperature anisotropy data provided by the Wilkinson Microwave Anistropy Probe (WMAP) combined with other cosmological probes, in particular the 2dF and SDSS galaxy redshift surveys. | $\Omega_M = 0.27$ $\pm 0.04$ $\Omega_\Lambda = 0.72$ $\pm 0.05$ $w = -1$ $\pm 6\%$ |
| 2007- | Efforts to explain Dark Energy continue, including work on theories of modified gravity. Ambitious new large ground-based surveys and space missions are planned. | |

## Newton, Hooke and Einstein

Isaac Newton's three volumes of *Principia* [29] were published, in Latin, in July 1687 and immediately had a significant effect on the scientific community and made Newton famous at the age of forty-five.

In propositions 70 to 71 he formulates the inverse square law of gravitation whereby a point mass $m$ situated outside a sphere of mass $M$ is attracted towards the centre of the sphere with a force $F$ inversely proportional to the square of its distance $r$ from the centre:

$$F = -\frac{GMm}{r^2}. \qquad (1)$$

He goes on to show that the force acts as if all the mass is concentrated in the centre of the sphere and that the same law holds between two different spheres, which meant that this mathematical analysis could be applied to the actual problems of astronomy.

What is intriguing is that, having completed this discussion, Newton explores the consequences of a wide range of central force laws and comes to the conclusion that there is a *second* form for which spherically symmetric masses can be treated as if all the mass is located at the central point. That is when "the compounded force with which two spheres attract each other is as the distance between the centres of the spheres" (Newton, Proposition 77, Theorem 37). He comments in the *Scholium*

> I have now explained the two principal cases of attractions: when the centripetal forces decrease as the square of the ratio of the distances, or increase in a simple ratio of the distances, causing the bodies in both cases to revolve in conic sections, and composing spherical bodies whose centripetal forces observe the same law of increase or decrease in the recess from the centre as the forces of the particles themselves do; which is very remarkable.

Subrahmanyan Chandrasekhar, who rewrote a large part of Books I and III of the *Principia* in a style more accessible to contemporary scientists, notes that this is the only place where Newton allows himself an expression of surprise [30].

As far as we know Newton did not consider the superposition of the two forces, but a straightforward conjecture would be to consider the full force law due to mass $M$ as:

$$\frac{F}{m} = \ddot{r} = -\frac{GM}{r^2} + CMr, \qquad (2)$$

G is the gravitational constant and $C$ is an arbitrary constant. We later relate $CM$ to the cosmological constant.

It is possible that Newton came to this conclusion partly due to the influence of Robert Hooke (1635-1694), a brilliant, but perhaps now somewhat overlooked scientist, seven years older than Newton. Newton was not, according to contemporary accounts, an easy man to get along with, and he began a life long feud with Hooke when Hooke criticized his writings on optics in the early 1670s. The extra term in the force equation is of the same form as the law of elasticity that Hooke had discovered in 1660, which states that the extension produced in a spring is proportional to the load, i.e. $F = kx$. Most likely Newton's interest in force laws and orbits was further developed from a debate between him and Hooke on the path of a heavy body falling in the Earth, which Newton initially claimed would spiral to the centre. Hooke's public announcement of Newton's mistake and his letter to Newton did nothing to improve the relationship between the two men. When the manuscript of *Principia* was first presented to the Royal Society in 1686, Edmond Halley wrote to Newton telling him that "Hooke had some pretensions to the invention of the rule for the decrease of gravity being reciprocally as the squares of the distances from the centre" (letter from Halley to Hooke, May 22, 1686). Hooke expected Newton to acknowledge his contribution in the preface, but Newton wrote a curt letter back to Halley claiming that he had come to his conclusions independently and that Hooke was of no consequence. Whatever the real truth, the dispute with Hooke undoubtedly revived Newton's interest in gravitational attraction and planetary motions.

In Book III of the *Principia,* 'The System of the World,' Newton discusses real astronomical observations and here he abandons the linear force term, presumably because there is no discernable evidence for it, and concludes simply that: "The force of gravity towards the several equal particles of any body is inversely as the square of the distances of places from the particles."

The idea that the force decreases as the inverse square of the distance had in fact originated some time earlier. Johannus Scotus Erigena (c.800 – c.877) guessed that heaviness and lightness vary with distance from the Earth, and this theory was take up by Adelard of Bath (twelfth century), while the first recorded suggestion of an inverse square law was made about 1640 by Ismael Bullialdus (1605 – 1694). Newton, however, was almost certainly the first, as early as 1665 or 1666, to deduce the inverse-square law observationally. It seems that he put off publishing the calculations for twenty years because he didn't know how to justify the fact that he had treated the Earth as if its whole mass were concentrated at its centre. In a letter to Halley on 20 June 1686 he wrote:

> I never extended the duplicate proportion lower than to the superficies of the earth, and before a certain demonstration I found last year, have suspected it did not reach accurately enough down so low; and therefore in the doctrines of projectiles never used it nor considered the motions of heavens.

Chandrasekhar holds the view that Newton's reluctance (even after 1679) to pursue his dynamical investigations arose from his dissatisfaction at not being able to conclusively prove or disprove this proposition, on which the exactitude of his entire theory rests.

Although Newton's gravitational theory was highly successful, it appeared to be unable to explain certain 'anomalies' of planetary motion, such as the precession of the perihelion of Mercury, and furthermore many people felt a philosophical uneasiness at the idea of action at a distance. In the century after his death several 'laws of gravitation' were formulated to rival Newton's but J.D. North [31] believes it likely that few of these theories were meant as more than mathematical exercises. Pierre-Simon Laplace was the first to explicitly write down the general force law $F = Ar + B/r^2$.

In the *Scholium* following Proposition 78 of the *Principia*, Newton remarks that both the inverse square force and the linear force cause the bodies to revolve in conic sections. In Proposition 10 he proves that "a particle will describe an ellipse about its centre under a centripetal attraction proportional to the distance…or perhaps in a circle into which the ellipse may degenerate." In fact, $F \propto r^{-2}$ and $F \propto r$ are the *only* two cases which allow stable planetary orbits or classical atomic orbits, which is quite surprising. But whereas the force inside a spherical shell is zero in the case of the inverse square law, it varies smoothly across the boundary for the linear force term. In Table 2 we contrast the properties of these two forces and their sum, for a point mass $m$ at a distance $r$ from the centre of a spherical shell of mass $M_s$ and radius $R$.

|  | $F/m = -GM/r^2$ | $F/m = CMr$ | $F/m = -GM/r^2 + CMr$ |
|---|---|---|---|
| Force inside a spherical shell | 0 | $CM_s r$ | $CM_s r$ |
| Potential inside a spherical shell | constant | $-\dfrac{CM_s}{2}(r^2 + R^2)$ | $const - \dfrac{CM_s}{2}(r^2 + R^2)$ |
| Force outside a spherical shell | $-\dfrac{GM_s}{r^2}$ | $CM_s r$ | $-\dfrac{GM_s}{r^2} + CM_s r$ |
| Potential outside a spherical shell | $-\dfrac{GM_s}{r}$ | $-\dfrac{CM_s}{2}(r^2 + R^2)$ | $-\dfrac{GM_s}{r} - \dfrac{CM_s}{2}(r^2 + R^2)$ |
| Shape of orbit caused by force[1] | any conic section | ellipse or circle | rosette |

**Table 2.** Contrasting $1/r^2$ and $r$ forces. The potential-force pairs $F = -\nabla \phi$ are per unit mass $m$ and the shell is characterized by mass $M_s$ and radius $R$. $C$ is an arbitrary constant.

It is uncertain to what extent Einstein owed a debt to previous theories of gravitation, but there is a remarkable similarity between Einstein's introduction of his cosmological constant, $\Lambda$, and the extra term in Newton's force law that was later ignored. If $\Lambda$ is included in the field equations, such that $G_{\mu\nu} - \Lambda g_{\mu\nu} = 8\pi G T_{\mu\nu}$, the Friedmann solution in the limit of weak gravity is written (assuming $p = 0$)

$$\frac{\ddot{a}}{a} = -\frac{4\pi G \rho}{3} + \frac{\Lambda}{3} \tag{3}$$

where $a$ is the scale factor. When applied to a sphere of radius $r$ and mass $M$, since density is mass/volume we have

$$\ddot{r} = -\frac{GM}{r^2} + \frac{\Lambda}{3} r \tag{4}$$

Comparing the above with equation (2) we see that they have exactly the same form.

## Einstein and Mach

In 1917 it was generally believed that the entire universe consisted of the Milky Way and the idea that it was static and unchanging was taken for granted. When Einstein introduced the cosmological constant in a paper entitled 'Cosmological Considerations on the General Theory of Relativity' he wrote in his conclusion that it was "for the purpose of making possible a quasi-static distribution of matter, as required by the fact of the small velocities of the stars" [32]. The paper makes no explicit mention of Ernst Mach, but it is clear that Einstein's conception of the universe, and thus his introduction of $\Lambda$, is not only strongly influenced by the status quo, but also by Machian ideas.

---

[1] The solutions are only exact using Newtonian mechanics. If we use general relativity to define the forces, the orbits will gradually precess.

Mach's major work was *The Science of Mechanics,* first published in 1883, and it made a "deep and persisting impression" on Einstein when he first read it as a student. The book was best known for its discussion of *Principia* and in particular a critique of Newton's concepts of absolute space and absolute motion. Mach analysed Newton's famous rotating bucket experiment and stated "For me, only relative motions exist and I can see, in this regard, no distinction between rotation and translation." His ultimate aim was to eliminate all metaphysical ideas from science, believing that "nothing is real except the perceptions, and all natural science is ultimately an economic adaptation of our ideas to our perceptions."

Whereas Newton defined a group of so-called 'inertial frames' that were at rest or in a state of uniform motion with respect to absolute space, Mach's inertial frames were determined relative to the fixed stars. He wondered "What would become of the law of inertia if the whole of the heavens began to move and the stars swarmed in confusion? How would we apply it then? How would it be expressed then? . . . Only in the case of the universe [do] we learn that *all bodies* [his italics] each with its share are of importance in the law of inertia" [33].

At the time, Einstein believed so strongly in the relativity of inertia that in 1918 he called it 'Mach's principle' and said it was a fundamental requirement of any satisfactory theory of gravitation [34]. The principle required that inertia should be fully and exclusively determined by matter, and since the metric $g_{\mu\nu}$ in the field equations determine the inertial action they should be impossible to determine in the complete absence of matter. "There can be no inertia relative to 'space,' but only an inertia of masses relative to one another" [32]. He defined his new fundamental constant $\Lambda$ in terms of the mass density $\rho$ of the universe, so that if $\Lambda$ is nonzero then the density must be nonzero. If $\rho = 0$ and there is no matter then there is no inertia because, it seemed, there could be no solution to the modified field equations. Unfortunately, just after the paper was published Willem de Sitter did find a solution to the modified field equations with $\rho = 0$, i.e. no matter in the universe at all, and then Alexander Friedmann and Georges Lemaitre found dynamic solutions to the original unmodified field equations. The final blow came when Edwin Hubble and Milton L. Humason discovered a rough proportionality of galaxy distances with their redshifts, and this was interpreted as evidence of an expanding universe. In 1954 Einstein wrote to a colleague "as a matter of fact, one should no longer speak of Mach's principle at all" [35].

## Problems with Newtonian cosmology

There is a problem with Newton's inverse square law if it is applied to an infinite universe with a nearly homogeneous matter distribution. In a Newtonian universe the gravitational force on a test body of unit mass is the resultant of the forces exerted by all the masses in the universe. Unfortunately, when this force is computed by an integration over all the masses, the integral fails to converge. It is surprising that Newton, a brilliant mathematician, did not see this, but perhaps he did not consider that the distribution of mass (i.e. stars) extended to infinity even though the space might. Newton probably viewed the universe as a finite system of stars and planets surrounded by infinite empty space. To explain the stability of the fixed stars he wrote in the *Principia*: "And lest the system of the fixed stars should, by their gravity, fall on each other, he [God] hath placed those systems at immense distances from one another."

Seeliger's papers of 1895 and 1896 investigated the difficulty in some depth and he came to the conclusion that it could be avoided by adding a tiny correction term to the inverse square law, whose effect would only become apparent at extremely large – cosmic – distances. The problem then was there were infinitely many possible modifications to Newton's law that would cause the integrals to converge, while still remaining compatible with observation, and there was no way to choose between them. Carl Neumann was able to come up with a unique modification of Newton's law in 1896, while in 1897 August Föppl and Lord Kelvin proposed refuting Newton's assumption of the universality of gravitation.

Given the attention Seeliger had brought to the matter, it is understandable that Einstein began his paper of 1917 with a re-analysis of the difficulties with Newtonian theory. His ultimate aim, however, was to provide additional justification for his introduction of a cosmological constant. He argued, as Seeliger had done, that Newtonian theory

> requires that the universe should have a kind of centre in which the density of the stars is a maximum, and that as we proceed outwards from this centre the group-density of the stars should diminish, until finally, at great distances, it is succeeded by an infinite region of emptiness. The stellar universe ought to be a finite island in the infinite ocean of space.

But Einstein went further and argued that even this island of stars would not be stable. Boltzmann's law of distribution for gas molecules would hold equally well for a cluster of stars, and this required that the cluster would gradually evaporate: "A vanishing of the density at infinity thus implies a vanishing of the density at the centre" [32]. Moreover, an island universe in flat spacetime violates Mach's idea of the relativity of inertia, which would require the universe to be homogenous and isotropic, with no arbitrarily isolated particles moving off to infinity.

If, however, Newton's inverse square law, written in the field form of Poisson's equation, is modified by the addition of an extra term $\Lambda\phi$, then the solution would correspond to an infinite extension of static space filled uniformly with matter. Here $\phi$ is the gravitational potential and "$\Lambda$ denotes a universal constant" and the modification is of the same form as Neumann's. Both the mean potential and mean density would remain constant to infinity.

Einstein writes that the addition of the cosmological constant "is perfectly analogous to the extension of Poisson's equation," thus implying that his introduction of $\Lambda$ into the field equations of GR has its roots in the failings, as he saw them, of Newtonian theory. In fact, it seems he was motivated mainly by his wish to find a solution to his field equations that was in accordance with Mach's principle and the prevalent orthodoxy of a static universe.

Interestingly, John Norton [36] has pointed out that Einstein's argument for the introduction of an extra term was based on a slightly flawed analysis of Newtonian cosmology. Franz Selety [37] showed that an island of stars in a flat, infinite universe would remain stable and not evaporate if its density diluted as $1/r^2$. And it is curious that Einstein, equally as brilliant a scientist as Newton, also neglected to think through the problems of the gravitational properties of an infinite matter distribution. Andrzej Trautman [38] is the first to show that Einstein's addition to the Poisson equation is not the correct non relativistic limit of general relativity with the cosmological term. The extra term is in fact simply $\Lambda$, such that

$$\nabla^2 \phi = 4\pi G \rho - \Lambda. \qquad (5)$$

John Barrow and Frank Tipler also remark on this [21]. The potential is then $\phi = -\frac{GM}{r} - \frac{\Lambda}{6}r^2$, where $\Lambda = 8\pi G \rho_\Lambda$.

## Is $\Lambda$ related to the total mass of the universe?

Considering $\Lambda$ as a Newtonian concept has a remarkable consequence which becomes clear if we compare equations (2) and (4). Identifying corresponding constants gives us $CM = \Lambda/3$. Using Gauss's Law, which describes the flux of a vector field through a surface, it can be shown that in the case of the linear force only, the mass corresponds to the mass of the entire universe [39]. The inverse square law due to a spherical distribution depends only on the mass within the spherical shell, which acts as if it is concentrated at the centre. But the linear force, due to any distribution whatsoever, will act as if the total mass of the universe is concentrated at its centre of mass. This is illustrated for the mass shell in Table 2. Thus we have

$$\Lambda \propto M_{tot}. \qquad (6)$$

In this interpretation $\Lambda$ is truly cosmological because all the mass in the universe contributes to it, if we consider a finite universe. Thus $\Lambda$ remains important even if the linear force is dominated by the inverse square force, as it would be in the vicinity of Earth, where the average density is far greater than that of the observable universe as a whole. Finding the value of $\Lambda$ could then perhaps lead us to an estimate of the universe's total mass. The relation could also possibly shed some light on the seemingly bizarre coincidence that the Dark Energy density $\rho_\Lambda$ is presently of the same order of magnitude as the mass density $\rho_M$. If $\Lambda$ is related to mass of the entire universe then maybe these two quantities are fundamentally connected.

●

Given all the current furore over Dark Energy, it is interesting that, 320 years ago, Newton discovered a term that might be related to it. He must have realised the implications of the $\Lambda$-like term and could not imagine, as Einstein could not, that the universe was expanding, or that the effects of the additional term would only become apparent at vast distances. It would be ironic if, after all the speculation and complications and theoretical ingenuity, observations show that Newton's original classical equations held the answer.

We thank Donald Lynden-Bell for stimulating discussions of Newtonian versions of Dark Energy. Many thanks also to Michela Massini, Shaun Thomas, Jochen Weller and the anonymous referee for helpful comments.